# Cryptanalysis of SDES via Evolutionary Computation Techniques

Poonam Garg
Information Technology and Management Dept.
Institute of Management Technology
Ghaziabad - India
pgarg@imt.edu

*Abstract*— *The cryptanalysis of simplified data encryption standard can be formulated as NP-Hard combinatorial problem. The goal of this paper is two fold. First we want to make a study about how evolutionary computation techniques can efficiently solve the NP-Hard combinatorial problem. For achieving this goal we test several evolutionary computation techniques like memetic algorithm, genetic algorithm and simulated annealing for the cryptanalysis of simplified data encryption standard problem (SDES). And second was a comparison between memetic algorithm, genetic algorithm and simulated annealing were made in order to investigate the performance for the cryptanalysis on SDES. The methods were tested and extensive computational results show that memetic algorithm performs better than genetic algorithms and simulated annealing for such type of NP-Hard combinatorial problem. This paper represents our first effort toward efficient memetic algorithm for the cryptanalysis of SDES*

I. INTRODUCTION

This paper proposes the cryptanalysis of simplified encryption standard algorithm using memetic and genetic algorithm. The cryptanalysis of simplified data encryption standard can be formulated as NP-Hard combinatorial problem. Solving such problems requires effort (e.g., time and/or memory requirement) which increases with the size of the problem. Techniques for solving combinatorial problems fall into two broad groups – traditional optimization techniques (*exact* algorithms) and non traditional optimization techniques (*approximate* algorithms). A traditional optimization technique guarantees that the optimal solution to the problem will be found. The traditional optimization techniques like branch and bound, simplex method, brute force search algorithm etc methodology is very inefficient for solving combinatorial problem because of their prohibitive complexity (time and memory requirement). Non traditional optimization techniques are employed in an attempt to find an adequate solution to the problem. A non traditional optimization technique - memetic algorithm, genetic algorithm, simulated annealing and tabu search were developed to provide a robust and efficient methodology for cryptanalysis. The aim of these techniques to find sufficient "good" solution efficiently with the characteristics of the problem, instead of the global optimum solution, and thus it also provides attractive alternative for the large scale applications. These nontraditional optimization techniques demonstrate good potential when applied in the field of cryptanalysis and few relevant studies have been recently reported.

In 1993 Spillman [16] for the first time presented a genetic algorithm approach for the cryptanalysis of substitution cipher using genetic algorithm. He has explored the possibility of random type search to discover the key (or key space) for a simple substitution cipher. In 1993, Spillman [17], also successfully applied a genetic algorithm approach for the cryptanalysts of a knapsack cipher. It is based on the application of a directed random search algorithm called a genetic algorithm. It is shown that such a algorithm could be used to easily compromise even high density knapsack ciphers. In 1997 Kolodziejczyk [11] presented the application of genetic algorithm in cryptanalysis of knapsack cipher .In 1999 Yaseen [19] presented a genetic algorithm for the cryptanalysis of Chor-Rivest knapsack public key cryptosystem. In this paper he developed a genetic algorithm as a method for Cryptanalyzing the Chor-Rivest knapsack PKC. In 2003 Grundlingh [9] presented an attack on the simple cryptographic cipher using genetic algorithm. In 2005 Garg [2] has carried out interesting studies on the use of genetic algorithm & tabu search for the cryptanalysis of mono alphabetic substitution cipher. In 2006 Garg [3] applied an attack on transposition cipher using genetic algorithm, tabu Search & simulated annealing. In 2006 Garg [4] studied that the efficiency of genetic algorithm attack on knapsack cipher can be improved with variation of initial entry parameters. In 2006 Garg[5] studied the use of genetic algorithm to break a simplified data encryption standard algorithm (SDES). In 2006

Garg[6] explored the use of memetic algorithm to break a simplified data encryption standard algorithm (SDES).

## II. THE SDES ALGORITHM DESCRIPTION

The SDES [18] encryption algorithm takes an 8-bit block of plaintext and a 10-bit key as input and produces an 8-bit block of cipher text as output. The decryption algorithm takes an 8-bit block of ciphertext and the same 10-bit key used as input to produce the original 8-bit block of plaintext. The encryption algorithm involves five functions; an initial permutation (IP), a complex function called $f_K$ which involves both permutation and substitution operations and depends on a key input; a simple permutation function that switches (SW) the two halves of the data; the function $f_K$ again, and a permutation function that is the inverse of the initial permutation ($IP^{-1}$). The function $f_K$ takes as input the data passing through the encryption algorithm and an 8-bit key. Consider a 10-bit key from which two 8-bit sub keys are generated. In this case, the key is first subjected to a permutation P10= [3 5 2 7 4 10 1 9 8 6], then a shift operation is performed. The numbers in the array represent the value of that bit in the original 10-bit key. The output of the shift operation then passes through a permutation function that produces an 8-bit output P8=[6 3 7 4 8 5 10 9] for the first sub key (K1). The output of the shift operation also feeds into another shift and another instance of P8 to produce subkey K2. In the second all bit strings, the leftmost position corresponds to the first bit. The block schematic of the SDES algorithm is shown in Figure 1.

Encryption involves the sequential application of five functions:

1. Initial and final permutation (IP).
   The input to the algorithm is an 8-bit block of plaintext, which we first permute using the IP function IP= [2 6 3 1 4 8 5 7]. This retains all 8-bits of the plaintext but mixes them up. At the end of the algorithm, the inverse permutation is applied; the inverse permutation is done by applying, $IP^{-1}$ = [4 1 3 5 7 2 8 6] where we have $IP^{-1}$ (IP(X)) =X.

2. The function $f_K$, which is the complex component of SDES, consists of a combination of permutation and substitution functions. The functions are given as follows. Let L, R be the left 4-bits and right 4-bits of the input, then,
   $f_K$ (L, R) = (L XOR f(R, key), R)
   where XOR is the exclusive-OR operation and key is a sub-key. Computation of f(R, key) is done as follows.
   1. Apply expansion/permutation E/P= [4 1 2 3 2 3 4 1] to input 4-bits.
   2. Add the 8-bit key (XOR).
   3. Pass the left 4-bits through S-Box $S_0$ and the right 4-bits through S-Box $S_1$.
   4. Apply permutation P4 = [2 4 3 1].

The two S-boxes are defined as follows:

$S_0$ $\quad\quad\quad\quad S_1$

$\begin{pmatrix} 1 & 0 & 3 & 2 \\ 3 & 2 & 1 & 0 \\ 0 & 2 & 1 & 3 \\ 3 & 1 & 3 & 2 \end{pmatrix}$ $\quad\quad \begin{pmatrix} 0 & 1 & 2 & 3 \\ 2 & 0 & 1 & 3 \\ 3 & 0 & 1 & 0 \\ 2 & 1 & 0 & 3 \end{pmatrix}$

The S-boxes operate as follows: The first and fourth input bits are treated as 2-bit numbers that specify a row of the S-box and the second and third input bits specify a column of the S-box. The entry in that row and column in base 2 is the 2-bit output.

3. Since the function $f_K$ allows only the leftmost 4-bits of the input, the switch function (SW) interchanges the left and right 4-bits so that the second instance of $f_K$ operates on different 4-bits. In this second instance, the E/P, $S_0$, $S_1$ and P4 functions are the same as above but the key input is K2.

## III. OBJECTIVE OF THE STUDY

Cryptanalytic attack on SDES belongs to the class of NP-hard problem. Due to the constrained nature of the problem, this paper is looking for a new solution that improves the robustness against cryptanalytic attack with high effectiveness. The objective of the study is:

- To determine the efficiency and accuracy of evolutionary computation techniques like memetic algorithm, genetic algorithm and simulated annealing for the cryptanalysis of SDES.
- To compare the relative performance of memetic algorithm, genetic algorithm and simulated annealing.

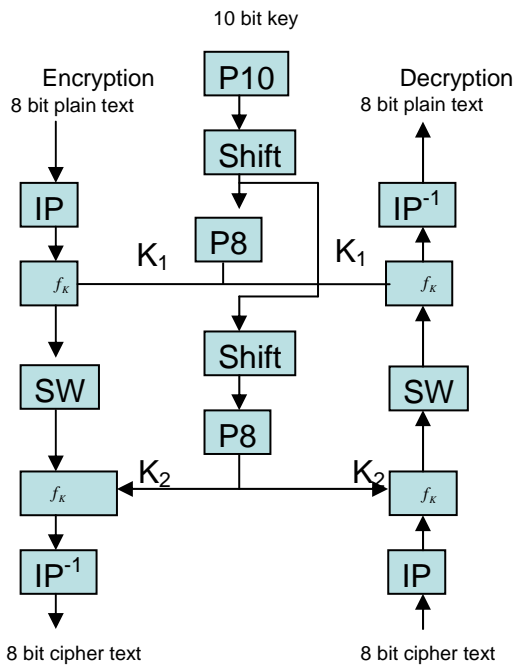

**Figure 1:** Simplified Data encryption algorithm

## IV. COST FUNCTION

The ability of directing the random search process of the genetic algorithm by selecting the fittest chromosomes among the population is the main characteristic of the algorithm. So the fitness function is the main factor of the algorithm. The choice of fitness measure depends entirely on the language characteristics must be known. The technique used by Nalini[13] to compare candidate key is to compare n-gram statistics of the decrypted message with those of the language (which are assumed known). Equation 1 is a general formula used to determine the suitability of a proposed key(k), here ,K is known as language Statistics i.e for English, [A,…….,Z_], D is the decrypted message statistics, and u/b/t are the unigram, bigram and trigram statistics. The values of α, β and γ allow assigning of different weights to each of the three n-gram types where α + β + γ =1.

$$C_k \approx \alpha . \sum_{i \in A} |K^u_{(i)} - D^u_{(i)}| + \beta . \sum_{i,j \in A} |K^b_{(i,j)} - D^b_{(i,j)}| + \gamma . \sum_{i,j,k \in A} |K^t_{(i,j,k)} - D^t_{(i,j,k)}| \quad (1)$$

When trigram statistics are used, the complexity of equation (1) is O(P3) where P is the alphabet size. So it is an expensive task to calculate the trigram statistics. Hence we will use assessment function based on bigram statistics only. Equation 1 is used as fitness function for genetic algorithm attack.

## V. METHODOLOGY

### A. Genetic algorithm approach

The genetic algorithm is based upon Darwinian evolution theory. The genetic algorithm is modeled on a relatively simple interpretation of the evolutionary process; however, it has proven to a reliable and powerful optimization technique in a wide variety of applications. Holland [10] in 1975 was first proposed the use of genetic algorithms for problem solving. Goldberg [7] were also pioneers in the area of applying genetic processes to optimization. As an optimization technique, genetic algorithm simultaneously examines and manipulates a set of possible solution. Over the past twenty years numerous application and adaptation of genetic algorithms have appeared in the literature. During each iteration of the algorithm, the processes of selection, reproduction and mutation each take place in order to produce the next generation of solution. Genetic Algorithm begins with a randomly selected population of chromosomes represented by strings. The GA uses the current population of strings to create a new population such that the strings in the new generation are on average better than those in current population (the selection depends on their fitness value). The selection process determines which string in the current will be used to create the next generation. The crossover process determines the actual form of the string in the next generation. Here two of the selected parents are paired. A fixed small mutation probability is set at the start of the algorithm. This crossover and mutation processes ensures that the GA can explore new features that may not be in the population yet. It makes the entire search space reachable, despite the finite population size. Figure 2 shows the generic implementation of genetic algorithm.

1. Encode solution space
2. (a) Set pop_size, max_gen, gen=0
   (b) set cross_rate, mutate_rate;
3. initialize population
4. while max_gen $\geq$ gen
    evaluate fitness
    for (i=1 to pop_size)
        select (mate1,mate2)
        if (rnd(0,1)$\leq$ cross_rate)
            child = crossover(mate1,mate2)
        if (rnd(0,1)$\leq$ mutate_rate)
            child = mutation();
        repair child if necessary
    end for
    Add offspring to new generation
    Gen=gen+1
    End while
5. return best chromosomes

**Figure 2 :** A generic genetic algorithm

### B. Memetic algorithm approach

The genetic algorithm is not well suited for fine-tuning structures which are close to optimal solution[7]. The memetic algorithms [15] can be viewed as a marriage between a population-based global technique and a local search made by each *of* the individuals. They are a special kind of genetic algorithms with a local hill climbing. Like genetic algorithms, memetic Algorithms are a population-based approach. They have shown that they are orders of magnitude faster than traditional *genetic Algorithms* for some problem domains. In a memetic algorithm the population is initialized at random or using a heuristic. Then, each individual makes local search to improve its fitness. To form a new population for the next generation, higher quality individuals are selected. The selection phase is identical inform *to* that used in the classical genetic algorithm selection phase. Once two parents have been selected, their chromosomes are combined and the classical operators of crossover are applied to generate new individuals. The latter are enhanced using a local search technique. The role of local search in memetic algorithms is to locate the local optimum more efficiently then the genetic algorithm. Figure 3 explains the generic implementation of memetic algorithm.

1. Encode solution space
2. (a) set pop_size, max_gen, gen=0;
   (b) set cross_rate, mutate_rate;
3. initialize population
4. while(gen < gensize)
      Apply generic GA
      Apply local search
   end while
   Apply final local search to best chromosome

**Figure 3:** The memetic algorithm

*1) Hill climbing local search algorithm*

The hill climbing search algorithm is a local search and is shown in figure 4. It is simply a loop that continuously moves in the direction of increasing quality value[15]

While (termination condition ins not satisfied) do
      New solution ← neighbors(best solution);
      If new solution is better then actual solution then
            Best solution ← actual solution
      End if
End while

**Figure 4 :** The Hill climbing local search algorithm

*C. Simulated Annealing approach*

In 1983 Kirkpatrick [12] proposed an algorithm which is based on the analogy between the annealing of solids and the problem of solving combinatorial optimization problems. An attack on the transposition cipher using simulated annealing is described in Figure 5.

1. Set the initial temperature, $T^{(0)}$.
2. Generate an initial solution - arbitrarily set to the identity transformation (could be randomly generated or otherwise).
3. Evaluate the cost function for the initial solution. Call this $C^{(0)}$.
4. For temperate *T* do many (eg., 100 × *M*) times:
      Generate a new solution by modifying the current one in some manner
      Evaluate the cost function for the newly proposed solution.
      Consult the Metropolis function to decide whether or not the newly proposed solution will be accepted.
      If accepted, update the current solution and its associated cost.
      If the number of accepted transitions for temperature *T* exceeds some limit (eg. 10 × *M*) then jump to Step 5.
5. If the number of accepted transitions for temperature *T* was zero then stop (return the current solution as the best), otherwise reduce the temperature (eg. $T^{(i+1)} = T^i \times 0.95$ and return to step 4

**Figure 5 :** *A simulated annealing*

VI. RESULT & DISCUSSIONS

In this section a number of experiments are carried out which outlines the effectiveness of all the three algorithms described above. The purpose of these experiments is to compare the performance of memetic algorithm, genetic algorithm and simulated annealing for the cryptanalysis of simplified SDE algorithm. The experiments were conducted on Pentium IV using 'C' language. Experimental results obtained from these algorithms were generated with 100 runs per data point e.g. ten different messages were created for both the algorithms and each algorithm was run 10 times per message. The best result for each message was averaged to produce data point.

For each algorithm there are number of different parameters which need to vary to "fine-tune" the optimization process. For the memetic algorithm, the population size was set to 10; the probabilities for crossover and mutation were both 0.5 for all the test problems because it was the best configuration found empirically for the memetic algorithm. For the genetic algorithm, the population size was set to 100, the probability for crossover was 0.95, and the probability for mutation was 0.05 for all test problems as it was the best configuration found empirically for the genetic algorithm. For simulated annealing the number of variables in a problem of this type can be enormous, making determination of an optimal solution impossible. Simulated annealing scans a small area of the solution space in the search for the global minimum. We did not use the same configuration for the memetic algorithm and the genetic algorithm because it would be disadvantageous to either of them if the other party's best configuration is used.

TABLE 1 : COMPARISON OF MEMETIC ALGORITHM, GENETIC ALGORITHM AND SIMULATED ANNEALING

| Amount of Cipher text | Memetic Algorithm | | | Genetic Algorithm | | | Simulated Annealing | | |
|---|---|---|---|---|---|---|---|---|---|
| | TIME (M) | Std. devi. | Number of bit matched in the key (N) | TIME (M) | std. devi. | Number of bit matched in the key (N) | TIME (M) | std. devi. | Number of bit matched in the key (N) |
| 100 | 5.1 | 4.70 | 8 | 2.62 | 4.82 | 6 | 2.41 | 4.5 | 8 |
| 200 | 14 | 3.40 | 6 | 4.5 | 6.13 | 6 | 2.61 | 3.53 | 6 |
| 300 | 15.3 | 2.72 | 5 | 2.13 | 6.01 | 4 | 2.33 | 3.31 | 4 |
| 400 | 12.5 | 2.27 | 7 | 2.35 | 4.61 | 6 | 2.36 | 3.43 | 7 |
| 500 | 10 | 2.16 | 6 | 2.52 | 4.61 | 6 | 2.68 | 3.41 | 6 |
| 600 | 5.5 | 1.86 | 8 | 2.07 | 4.37 | 7 | 2..30 | 3.77 | 7.4 |
| 700 | 3.05 | 1.73 | 7 | 4.07 | 4.42 | 6 | 3.89 | 2.53 | 6 |
| 800 | 2.85 | 1.59 | 8 | 2.4 | 3.39 | 8 | 2.6 | 2.49 | 8 |
| 900 | 2.24 | 1.56 | 9 | 2.53 | 2.23 | 6 | 2.45 | 2.16 | 8 |
| 1000 | 2.14 | 1.49 | 9.17 | 2.17 | 2.20 | 7 | 2.34 | 2.12 | 8.3 |

Table 1 depicts the results of the memetic algorithm along with a comparison of genetic algorithm and simulated annealing. This table basically compares the average number of key elements (out of 10) correctly recovered versus the amount of cipher text and the computation time to recover the keys from the search space. The table shows results for amounts of cipher text ranging from 100 to 1000 characters.

From figure 6, the first point to note is that the numbers of keys obtained from both the algorithms are acceptable. From Table 1, it can be seen that the standard deviation values for memetic algorithm is less than genetic algorithm and simulated annealing, this shows that memetic algorithm has a less variance in its results. So statistically, it can be proved that the performance of memetic algorithm approach is slightly superior to genetic algorithm and simulated annealing for the cryptanalysis of SDES.

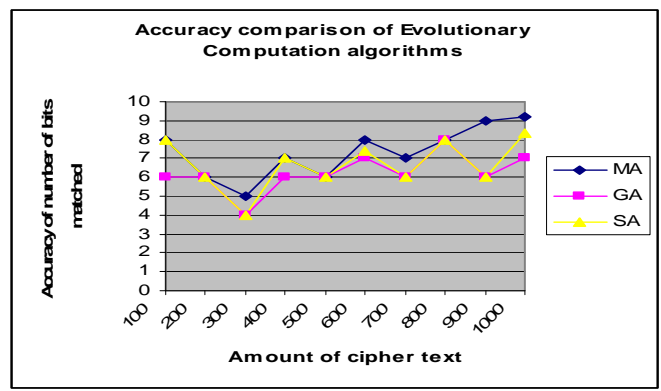

**Figure 6** : The Accuracy comparison of evolutionary computation algorithms

This may be because when the search technique is incorporated in algorithm then the solution space in better searched. According to the experimental results we can conclude that the local heuristic play an important role in memetic algorithm. Also we can say that including a high quality heuristic solution can help the memetic algorithm to improve its performance by reducing the likelihood of its premature convergence.

Comparing the running time of these algorithms, we found that genetic algorithm and simulated annealing is not sensitive to the amount of cipher text. Figure 7 clearly shows that the running time of memetic algorithm is severely reduced as we are increasing the amount of cipher text whereas results suggest that the genetic algorithm and simulated annealing is unaffected. Genetic algorithm and simulated annealing can be seen to be the most efficient algorithm as almost same keys is achieved in shorter time. In contrast memetic algorithm is more sensitive to amount of cipher text, for a large amount of cipher text the memetic algorithm can be seen outperform Genetic algorithm and simulated annealing . It means a small amount of cipher text provides an insufficient search space, which memetic algorithms perform poorly. And memetic algorithm has a slower evolution then genetic algorithm and simulated annealing because it has a high local search cost. However, a large amount of cipher text is having the large search space, possibly resulting improvement in case of memetic algorithm.

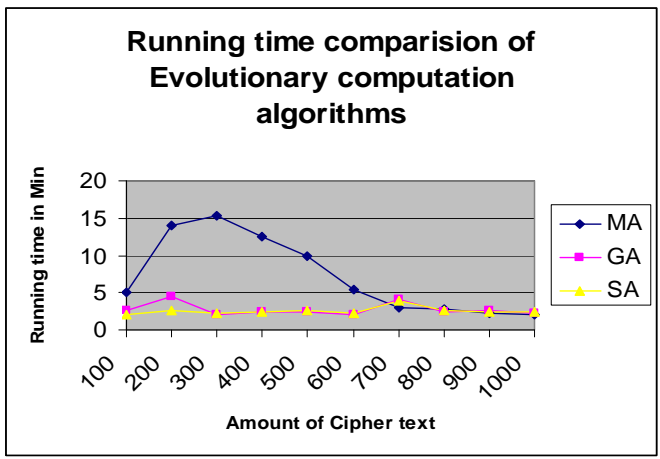

**Figure 7**: The running time comparison of evolutionary computation algorithms

## VII. CONCLUSION

In this paper we have presented a memetic algorithm, genetic algorithm and simulated annealing approach for the cryptanalysis of simplified data encryption standard algorithm – A challenging optimization problem in NP-Hard combinatorial problem. Our objective is to determine the performance of memetic algorithm in comparison with genetic algorithm and simulated annealing for the cryptanalysis of SDES. The first performance comparison was made on the average number of key elements (out of 10) correctly recovered versus the amount of ciphertext. Our experimental result shows that memetic algorithm is slightly superior for finding the number of keys accurately in comparison of genetic algorithm because search technique is incorporated in the algorithm and the solution space is better searched. The second comparison was made upon the computation time for recovering the keys from the search space. From the extensive experiments, it was found that simulated annealing algorithm can be seen to be the most efficient algorithm as almost same keys is achieved in shorter time but in contrast for a large amount of cipher text the memetic algorithm can be seen outperform simulated algorithm. Result indicates that memetic algorithm is extremely powerful technique for the cryptanalysis of SDES.

REFERENCES

[1] Davis, L. , "Handbook of Genetic Algorithms", Van Nostrand Reinhold, New York ,1991.

[2] Garg Poonam, Sherry A.M., Genetic algorithm & tabu search attack on the monoalphabetic substitution cipher, Paradime Vol IX no.1, January-June 2005,pg 106-109

[3] Garg Poonam, Shastri Aditya, Agarwal D.C, Genetic Algorithm, Tabu Search & Simulated annealing Attack on Transposition Cipher ,proceeding of Third AIMS International conference on management at IIMA – 2006, pg 983-989

[4] Garg Poonam, Shastri Aditya, An improved cryptanalytic attack on knapsack cipher using genetic algorithm approach, International journal of information technology, volume 3 number 3 2006, ISSN 1305-2403, 145-152

[5] Garg Poonam, Genetic Algorithm Attack on Simplified Data Encryption Standard Algorithm, International journal Research in Computing Science, ISSN 1870-4069, 2006.

[6] Garg Poonam, Memetic Algorithm Attack on Simplified Data Encryption Standard Algorithm, proceeding of International Conference on Data Management, February 2008, pg 1097-1108 .

[7] Goldberg, D.E., "Genetic Algorithms in Search, Optimization and Machine Learning", Addison-Wesley, Reading, 1989.

[8] Giddy J. P and Safavi-Naini R., Automated cryptanalysis of transposition ciphers, The Computer Journal, Vol 37, No. 5, 1994.

[9] Gr¨undlingh, W. & van Vuuren, J. H., Using Genetic Algorithms to Break a Simple Cryptographic Cipher, Retrieved March 31, 2003 from

http://dip.sun.ac.za/˜vuuren/abstracts/abstr genetic.htm, submitted 2002.

[10] Holland, J., "Adaptation in Natural and Artificial Systems", University of Michigan Press, Ann Arbor, 1975.

[11] Kolodziejczyk, J., Miller, J., & Phillips, P. ,The application of genetic algorithm in cryptoanalysis of knapsack cipher, In Krasnoproshin, V., Soldek, J., 1997

[12] Kirkpatrick S., Gelatt C. D., Jr., and Vecchi M. P., "Optimization by simulated annealing", *Science,* 220(4598):671–680, 1983.

[13] Nalini, Cryptanalysis of Simplified data encryption standard via Optimization heuristics, International Journal of Computer Sciences and network security, vol 6, No 1B, Jan 2006

[14] P. Men and B. Freisleben, "Memetic Algorithms for the Traveling Salesman Problem," Complex Systems, 13(4):297-345. 2001.

[15] P, Moscato, "on evolution, scorch, optimization. genetic algorithms and martial arts: toward memetic olgorithms", Technical report, California, 1989.

[16] Spillman et. al., Use of a genetic algorithm in the cryptanalysis of simple substitution ciphers, Cryptologia, 17(1):187-201, April 1993.

[17] Spillman R.,Cryptanalysis of knapsack ciphers using genetic algorithms. Cryptologia, 17(4):367–377, October 1993.

[18] Schaefer E, A simplified data encryption standard algorithm, Cryptologia, Vol 20, No 1, 77-84, 1996.

[19] Yaseen, I.F.T., & Sahasrabuddhe, H.V. (1999), A genetic algorithm for the Cryptanalysis of Chor-Rivest knapsack public key cryptosystem (PKC), In Proceedings of Third International Conference on Computational Intelligence and Multimedia Applications, pp. 81-85, 1999

AUTHOR PROFILE

**Dr. Poonam Garg** is an Associate Professor and Chairperson IT Infrastructure at IMT, Ghaziabad, India. Dr. Garg received her M.C.A. degree in 1991 and Ph. D. Degree in 2006 in Cryptology from Banasthali University, Rajasthan, India. Dr. Garg's current research interests are in the area of developing heuristics and meta-heuristics particularly Genetic Algorithm, Tabu Search and Simulated Annealing based meta-heuristic for various optimization problem such as cryptanalysis of various encryption algorithms, scheduling and project management

Dr. Garg has 17+ years of experience in teaching, research and consulting. She has conducted large number of Management Development Programs for middle and senior level management in public and private sectors. She has about 25 papers in different journals and conferences, and has four edited books. She has served many International Conferences as a conference cochair, conference track chair and members of program committee.